\documentstyle [11pt, epsfig]{article}
\begin{document}
 \def\la{\langle}
 \def\ra{\rangle}
\hfill {WM-05-119}

\hfill {\today}

\vskip 1in   \baselineskip 24pt

{
\Large
   \bigskip
   \centerline{3-3-1 Models with Unique Lepton Generations}
 }
\def\bar{\overline}

\centerline{David L. Anderson\footnote{Email: dlande@wm.edu} and Marc 
Sher\footnote{Email: sher@physics.wm.edu}}
\centerline {\it Particle Theory Group}
\centerline {\it Department of Physics}
\centerline {\it College of William \& Mary, Williamsburg, VA 23187, 
USA}
\bigskip

{\narrower\narrower
We study previously unconsidered 3-3-1 models which are characterized by each lepton generation having a
different representation under the gauge group.
Flavor-changing neutral currents in the lepton sector occur in these models.
To satisfy constraints on $\mu \rightarrow 3e$ decays, the $Z'$ must be heavier than 2 to 40 TeV, depending on 
the model and assignments of the leptons.
These models can result in very unusual Higgs decay modes. 
In most cases the $\mu \tau$ decay state is large (in one case, it is the dominant mode), and in one case,
the $\Phi \rightarrow s\bar{s}$ rate dominates. }

\newpage
\section{Introduction}\label{sec:intro}

An interesting extension of the standard model is based on the gauge 
group $SU(3)_{c}\times SU(3)_{L}\times U(1)$ ($331$).  In the 
original, minimal version of the model\cite{frampton,pisano}, the 
leptons are put into antitriplets of $SU(3)_{L}$, two generations of 
quarks are put into triplets and the third generation of quarks is 
put into an antitriplet.  With this structure, the anomalies will all 
cancel if and only if the number of generations is a multiple of 
three.  
The model has an automatic Peccei-Quinn symmetry\cite{pq,dias}, and 
the fact that one quark family has different quantum numbers than the other
two may explain the heavy top quark mass\cite{fram}.   An unusual 
feature of this model is that $\sin^{2}\theta_{W}$ must be less than 
$1/4$.  Since it is an increasing function of $q^{2}$, the scale of 
$SU(3)_{L}$ breaking must be relatively low, and cannot arbitrarily be
moved up to a high scale.

This minimal model contains doubly charged gauge fields (bileptons) 
as 
well as isosinglet quarks with exotic charges.   The phenomenology of 
these models is very rich and has been the subject of extensive 
study\cite{lots}.    A completely different class of models was 
proposed in Refs. \cite{montero,foot}, in which the embedding of the 
charge operator into $SU(3)_{L}$ is different.  In these models, 
there 
are no exotic charges for the quarks, and the gauge bosons are all 
either neutral or singly-charged.   In all of these models, one still 
treats the lepton generations identically, and treats one quark generation
differently than the other two.   A comprehensive review of the 
gauge, fermion and scalar sectors of all of these models can be found 
in Refs. \cite{ponce1,ponce2}.

In Ref. \cite{ponce1}, a detailed analysis of the anomalies in 331 
models showed that there are two anomaly free sets of fermion 
representations in which the lepton generations are {\bf all} treated 
differently.  The phenomenology of these models has never been 
studied in the literature.   With leptons in different 
representations, one might expect lepton-flavor changing neutral 
processes.

In this paper, we discuss the phenomenology of these two models.  In 
Section 2, the various 331 models are presented, as well as the 
possible representations for fermions in these models.  A set of 
anomaly free models will be found, and it will be noted that two of 
them have very different representations for the lepton families.
In Section 3, we will consider the scalar sector of these ``unique
lepton generation'' models, 
and in Section 4 will 
present the mass matrices for the leptons, look at the possible 
variations that can occur, and find the Yukawa couplings to the 
scalars.  The phenomenology of lepton-number violating $\mu$ and 
$\tau$ decays will be discussed in Section 5, and for Higgs decays in 
Section 6.   Our most interesting result will be that many of these models
have fairly large branching ratios for the Higgs boson decaying 
into a muon and a tau, and in one model it may be the dominant decay.
In Section 7, we will examine lepton-number violation due to gauge 
boson exchange, and the resulting bounds on the gauge boson masses.
Finally, in Section 8 we present our conclusions.
\section{Models}

As discussed in Ref. \cite{ponce1}, if one assumes that the isospin 
$SU(2)_{L}$ of the standard model is entirely embedded in 
$SU(3)_{L}$, 
then all models can be characterized by the charge operator
\begin{equation}
    Q = T_{3L} + {2\over \sqrt{3}}bT_{8L} + X I_{3}
    \end{equation}
where $I_{3}$ is the unit matrix and $T_{iL}=\lambda_{iL}/2$, where 
the $\lambda_{iL}$ are the Gell-Mann matrices.  $X$ is fixed by 
anomaly cancellation and the coefficient can be absorbed in the 
hypercharge definition.  Different models are characterized by 
different values of $b$.

In the original Frampton, Pisano, and Pleitez~\cite{frampton, pisano} model, $b=3/2$, leading
to doubly-charged gauge bosons and fermions with exotic charges.  The 
fermion representations, with the $SU(3)\times U(1)$ quantum numbers, 
are
\begin{equation}
    L_{i}=\pmatrix{e_{i}\cr \nu_{i}\cr e^{c}_{i}\cr}: (3^{*},0)
\end{equation}
for the leptons ($i=1,2,3$) and
\begin{equation}
    Q_{1,2}=\pmatrix{u\cr d\cr D\cr}, \pmatrix{c\cr s\cr S\cr}:  
(3,-1/3)
\end{equation}
\begin{equation}
    Q_{3}=\pmatrix{b\cr t\cr T\cr}: (3^{*},2/3)\end{equation}
with all of the quark conjugate fields being isosinglets.  $D,S,T$ 
are quarks with charges given by $-4/3, -4/3, 5/3$.

A simple variant of this model\cite{tully} changes the lepton 
structure by replacing the $e^{c}$ with a heavy lepton $E^{+}$ and 
adding $e^{c}$ and $E^{-}$ as singlets.

If one wishes to avoid exotic electric charges, one must choose 
$b=1/2$.   In that case, the fermion structure is very different.  
Following \cite{ponce1}, we can find six sets of fermions, 
which contain the antiparticles of all charged particles.  The first 
four are leptons and the last two are quarks.   Noting 
$e_{i}, d_{i}, u_{i}$ as standard model fermions, and 
$E_{i},D_{i},U_{i}$ as exotic fermions, the four sets of leptons are
\begin{equation}
L_{1} = \pmatrix{\nu_{i}\cr e_{i}^{-}\cr E_{i}^{-}\cr}; e_{i}^{+}; 
E_{i}^{+}
\label{eq:L1}
\end{equation}
with $SU(3)\times U(1)$ quantum numbers $(3,-2/3),(1,1),(1,1)$,
\begin{equation}
L_{2} = \pmatrix{e_{i}^{-}\cr \nu_{i}\cr N_{i}^{0}\cr}; e_{i}^{+}
\label{eq:L2}
\end{equation}
with $SU(3)\times U(1)$ quantum numbers $(3^{*},-1/3),(1,1)$, and 
$N^{0}_{i}$ is a heavy neutrino,
\begin{equation}
L_{3} = \pmatrix{e_{i}^{-}\cr \nu_{i}\cr N_{1}^{0}\cr}; 
\pmatrix{E_{i}^{-}\cr N_{2}^{0}\cr 
N_{3}^{0}\cr};\pmatrix{N_{4}^{0}\cr E_{i}^{+}\cr e_{i}^{+}\cr}
\label{eq:L3}
\end{equation}
with $SU(3)\times U(1)$ quantum numbers 
$(3^{*},-1/3),(3^{*},-1/3),(3^{*},2/3)$, and there are four heavy 
neutrino states (some may be conjugates of another), and
\begin{equation}
L_{4} = \pmatrix{\nu_{i}\cr e^{-}_{i}\cr E_{1i}^{-}\cr}; 
\pmatrix{E_{2i}^{+}\cr N_{1}^{0}\cr 
N_{2}^{0}\cr};\pmatrix{N_{3}^{0}\cr E_{2i}^{-}\cr
E_{3i}^{-}\cr};e_{i}^{+};E_{1i}^{+};E_{3i}^{+}
\label{eq:L4}
\end{equation}
with $SU(3)\times U(1)$ quantum numbers 
$(3,-2/3),(3,1/3),(3,-2/3),(1,1),(1,1),(1,1)$.

The two sets of quarks are
\begin{equation}
Q_{1}=\pmatrix{d_{i}\cr u_{i}\cr 
U_{i}\cr};d^{c}_{i};u^{c}_{i};U_{i}^{c}
\end{equation}
with $SU(3)\times U(1)$ quantum numbers 
$(3^{*},1/3),(1,1/3),(1,-2/3),(1,-2/3)$, and 
\begin{equation}
Q_{2}=\pmatrix{u_{i}\cr d_{i}\cr 
D_{i}\cr};u^{c}_{i};d^{c}_{i};D_{i}^{c}
\end{equation}
with $SU(3)\times U(1)$ quantum numbers 
$(3,0),(1,-2/3),(1,1/3),(1,1/3)$.

The anomalies for these six sets are\cite{ponce1} found in Table 1.
\begin{table}[h]
\begin{center}
\begin{tabular}
{||l||c|c|c|c|c|c||} \hline \hline
Anomalies & $L_1$ & $L_2$ & $L_3$ & $L_4$ & $Q_1$ & $Q_2$ \\ \hline \hline
$[SU(3)_c]^2U(1)_X$ & 0 & 0 & 0 & 0 & 0 & 0 \\
$[SU(3)_L]^2U(1)_X$ & -2/3 & -1/3 & 0 & -1 & 1 & 0 \\
$[grav]^2U(1)_X$ & 0 & 0 & 0 & 0 & 0 & 0 \\
$[U(1)_X]^3$ & 10/9 & 8/9 & 6/9 & 12/9 & -12/9 & -6/9 \\ \hline \hline
\end{tabular}
\end{center}
\caption{Anomalies for the Fermion Families}
\end{table}
With this table, anomaly-free models (without exotic charges) can be
constructed.  As noted in Ref. \cite{ponce1}, there are two 
one-family and eight three family models that are anomaly-free.  Of
the eight three-family models, four treat the lepton generations 
identically, two treat two of the lepton generations identically and
in two, the lepton generations are all different.  It is the latter 
two that will be the subject of this study.

Note that  one can easily see from Table 1 that there 
are only two one-family models.  The first consists 
of $Q_{2}+L_{3}$.  This structure is perhaps most familiar to grand 
unified model builders, since the $27$ fields are contained in the 
$27$-dimensional fundamental representation of $E_{6}$.  In addition 
to analyses of $E_{6}$ models, an analysis of this model, in the 
context of $331$ models, can be found in Refs. \cite{ponce1,ponce3}.

The second one-family structure is $Q_{1}+L_{4}$.  This model is
related to $SU(6)\times U(1)$ unified models, and is analyzed in 
Ref. \cite{ponce4}.  Note that both of these one-family models are 
simply triplicated to become three-family models.

There are two other three-family models in which all of the leptons 
are 
treated the same way (but now the quark generations are treated 
differently).  These were the first models analyzed once it 
was recognized that $331$ models without exotic charges (i.e. with 
$b=1/2$) could be constructed.  The first is $3L_{2}+Q_{1}+2Q_{2}$.  
As in the original $331$ models, one generation of quarks is treated
differently than the other two, and thus three families are needed to
cancel anomalies.  These were analyzed in Ref. \cite{foot}.  The 
second such model is $3L_{1}+2Q_{1}+Q_{2}$, which also requires three 
families for anomaly cancellation.  This model has been analyzed in 
Ref. \cite{ozer}.  

Two models involve simple replication of the two one-family models, 
but 
take two copies of the first one-family model and one copy of the 
second, or vice-versa, i.e. $2(Q_{2}+L_{3}) + (Q_{1}+L_{4})$ and 
$2(Q_{1}+L_{4}) + (Q_{2}+L_{3})$.  Since the lepton generations are 
not all different, we will not consider these models further, 
although 
they have not, to our knowledge, been studied.

The two models of interest treat all of the lepton generations 
differently.  They are Model A:  $L_{1}+L_{2}+L_{3}+Q_{1}+2Q_{2}$ and 
Model B: $L_{1}+L_{2}+L_{4}+2Q_{1}+Q_{2}$.  Note that each model has
two ``simple'' lepton families ($L_{1}$ and $L_{2}$ above), and one 
more complicated family.   We now analyze the phenomenology of these 
two models.  Note that one cannot determine which ($e,\mu,\tau$)
lepton belongs to which representation, and so we will consider all 
six possible permutations for each model.

\section{The Scalar Sector}

The scalar sector of 331 models has been extensively 
studied\cite{tully,diaz}.  Here, one can see a substantial advantage 
to $b=1/2$ models.  In the original $b=3/2$ models, the minimal Higgs 
sector consists of three $SU(3)_{L}$ triplets plus an $SU(3)_{L}$ 
sextet.  In the $b=1/2$ models, three triplets are sufficient.  One 
triplet breaks the $SU(3)_{L}\times U(1)$ gauge symmetry down to the 
standard model, and the other two are necessary to break the 
$SU(2)_{L}$ symmetry and to give the fermions 
mass.    A very comprehensive analysis of the scalar sector in all 
previously considered models can be found in Ref. \cite{diaz}.

Although the models we are considering are $b=1/2$ models, it is not a 
priori obvious that three triplets will suffice to give the leptons 
mass, since the different families have very different structure.
Our Model A has five charged leptons (the $e,\mu,\tau$ and two exotic 
leptons), and Model B has seven charged leptons (with four exotic 
leptons).  Fortunately, as will be seen in Section 4, three triplets 
will suffice to give the charged leptons mass.  We will not consider
neutrino masses in this study since the number of fields and the
various options (which exotic neutrinos correspond to which 
right-handed neutrinos, for example) will rule out any substantial 
predictive power.

The first stage of breaking from $SU(3)_{L}\times U(1)$ to 
$SU(2)\times U(1)$ is carried out by a triplet Higgs, $\Phi_{A}$, 
which is a $(3,1/3)$ under the $SU(3)_{L}\times U(1)$ group,
and its vev is given by
\begin{equation}
    \langle{\Phi_{A}}\rangle = \pmatrix{0\cr 0\cr V\cr}
    \end{equation}
Note that the second component of the triplet is neutral, and could 
also get a vev, but that can be removed by a gauge transformation.  
Five of the gauge bosons acquire masses of $O(V)$, while the remaining four
are massless at this stage.  One can easily see that this vev will 
give masses of $O(V)$ to the $U$ and $D$ exotic quarks, and in 
previously considered models, to the $E$ exotic leptons as well. 
These masses are phenomenologically constrained to be 
substantially larger than the electroweak scale.

The second stage of symmetry breaking requires two Higgs triplets,
$\Phi_{1}$ and $\Phi_{2}$ with quantum numbers $(3,-2/3)$ and
$(3,1/3)$ respectively.   If one only wished to break the gauge
symmetry, then one triplet would suffice.  However, giving mass to the
fermions requires a second doublet.  This is not too surprising, since 
the quark masses in the standard model necessitate a Higgs doublet $H$ 
and $i\tau_{2} H^{*}$ to give masses to the down and up quarks, 
respectively. In $SU(2)$ $\overline{2}=2$, but this does not apply
in $SU(3)$. Thus the low-energy theory is a two-doublet model.  The
vevs of these doublets are
\begin{equation}
    \langle{\Phi_{1}}\rangle=\pmatrix{v_{1}/\sqrt{2}\cr 0\cr 0\cr};\qquad 
    \langle{\Phi_{2}}\rangle=\pmatrix{0\cr v_{2}/\sqrt{2}\cr 0\cr}
    \end{equation}
where $v_{1}^{2}+v^{2}_{2} = (246\ {\rm GeV})^{2}$.  Note that the 
third component of $\Phi_{2}$ could acquire a nonzero vev, but this 
will not involve $SU(2)$ breaking and will be irrelevant.

\section{Yukawa Couplings}

 With the fermion representations discussed in Section 2 and the
 scalar representations discussed in Section 3, we can now write down
 the Yukawa couplings and mass matrices for the charged leptons.  Let
 us first write down the fermion representations more explicitly.

 For Model A, the fields, followed by their $SU(3)_{L}\times U(1)$
 quantum numbers, are (with the subscript $L$ understood)
 \begin{equation}
 \psi_{i}=\pmatrix{\nu_{i}\cr e_{i}\cr E_{i}\cr}, (3,-2/3);\quad  e^{c}_{i}, (1,1); \quad  E^{c}_{i}, (1,1) \end{equation}
 \vskip 0.5cm \begin{equation}
 \psi_{j}=\pmatrix{e_{j}\cr \nu_{j}\cr N^{o}_{j}\cr},   (\overline{3},-1/3);\quad e^{c}_{j}, (1,1) \end{equation}
 \vskip 0.5cm \begin{equation}
 \psi_{k}=\pmatrix{e_{k}\cr  \nu_{k}\cr N^{o}_{1k}\cr},    (\overline{3},-1/3); \quad \psi^{\prime}_{k}=\pmatrix{E_{k}\cr   
 N^{o}_{2k}\cr N^{o}_{3k}\cr},  (\overline{3},-1/3);\quad  \psi^{\prime\prime}_{k}=\pmatrix{N^{o}_{4k}\cr  E^{c}_{k}\cr e^{c}_{k}\cr},
  (\overline{3},2/3) \end{equation}
 where the $N^{o}$ could be a conjugate of either the $\nu$  or 
 another $N^{o}$, and the generation labels $i,j$ and $k$ are all distinct.
 Note that the model contains five charged leptons: the standard three
 plus two exotic leptons.
 
 For Model B, the fields are
 \begin{equation}
 \psi_{i}=\pmatrix{\nu_{i}\cr e_{i}\cr E_{i}\cr}, (3,-2/3);\quad  e^{c}_{i}, (1,1); \quad  E^{c}_{i}, (1,1) \end{equation}
 \vskip 0.5cm \begin{equation}
 \psi_{j}=\pmatrix{e_{j}\cr \nu_{j}\cr N^{o}_{j}\cr},   (\overline{3},-1/3);\quad e^{c}_{j}, (1,1) \end{equation}
 \vskip 0.5cm \begin{equation}
 \psi_{k}=\pmatrix{\nu_{k}\cr  e_{k}\cr E_{1k}\cr},    ({3},-2/3); \quad  \psi^{\prime}_{k}=\pmatrix{E^{c}_{2k}\cr   
 N^{o}_{1k}\cr  N^{o}_{2k}\cr},  ({3},1/3);\quad  \psi^{\prime\prime}_{k}=\pmatrix{N^{o}_{3k}\cr   E_{2k}\cr E_{3k}\cr}, ({3},-2/3);
 \quad e_{i}^{+};\quad E^{c}_{1k}; \quad E^{c}_{3k} \end{equation}
where the last three fields are singlets.   Note that this model has
seven charged leptons: the standard three plus four exotics.

From these representations, and the scalar fields (with their vevs)
in Section 3, we can write down the mass matrices for the charged 
leptons.  The mass matrix for Model A is $5\times 5$ and for Model B 
is $7\times 7$.  From these matrices, the Yukawa couplings to each 
scalar field can be trivially obtained by replacing the vev with the 
field.  The Yukawa couplings and full mass matrices are given in Appendix A.   If one takes
the limit in which $v_{1}=v_{2}=0$, then each of these matrices has 
three zero eigenvalues, indicating that the exotic leptons all get 
masses of $O(V)$.   Since $V$ must be 
large, we can take the limit as $V\rightarrow\infty$, and find 
the effective mass matrices for the three standard model leptons.  Note
that we do not know, a priori, which of the leptons is in the first, 
second, or third rows, so each model will have six permutations.

For Model A, we find that the mass matrix is of the form
\begin{equation}
    M_{A}={1\over\sqrt{2}}\pmatrix{h_{1}v_{2}&h_{2}v_{2}&0\cr h_{3}v_{1}&h_{4}v_{1}&h_{5}v_{2}\cr     h_{6}v_{1}&h_{7}v_{1}&h_{8}v_{2}\cr}
    \end{equation}
where the $h_{i}$ are constants.   The Yukawa coupling matrices are 
then
\begin{equation}
\pmatrix{0&0&0\cr h_{3}&h_{4}&0\cr h_{6}&h_{7}&0\cr}\Phi_{1}+\pmatrix{h_{1}&h_{2}&0\cr 0&0&h_{5}\cr 0&0&h_{8}\cr}\Phi_{2}.
\end{equation}

For Model B, the mass matrix is of the form
\begin{equation}
M_{B}={1\over\sqrt{2}}\pmatrix{h'_{1}v_{2}&h'_{2}v_{2}&h'_{3}v_{2}\cr h'_{4}v_{1}&h'_{5}v_{1}&h'_{6}v_{1}\cr h'_{7}v_{2}&h'_{8}v_{2}&h'_{9}v_{2}\cr}    
\end{equation}
and the Yukawa coupling matrices are \begin{equation}
\pmatrix{0&0&0\cr h'_{4}&h'_{5}&h'_{6}\cr 0&0&0\cr}\Phi_{1}+\pmatrix{h'_{1}&h'_{2}&h'_{3}\cr 0&0&0 \cr h'_{7}&h'_{8}&h'_{9}\cr}\Phi_{2}.
\end{equation}

These Yukawa coupling matrices are certainly unusual.  Note that 
diagonalizing the mass matrices will {\it not} diagonalize the Yukawa 
coupling matrices, and thus one will have lepton-flavor-changing 
neutral currents in the Higgs sector.  This is just the 
Glashow-Weinberg theorem\cite{gw}.  To determine the size of the 
lepton-flavor violation, one simply must diagonalize the mass matrix 
and read off the Yukawa coupling matrices in the diagonalized basis.

Unfortunately, such a procedure will not be useful.  The matrices 
have far too many free parameters.   Worse, in general fine-tuning will be needed.
We define ``fine-tuning'' as a situation in which several terms add 
together to give a term that is much smaller than any individual term.  In
general, fine-tuning will be needed to give the electron a small 
mass\footnote{There are trivial exceptions.  For example, in $M_{A}$, 
if $h_{1}$ is very small, and all off-diagonal terms vanish, then 
there is no fine-tuning (and no flavor-changing neutral currents).}, 
and it is unclear how this fine-tuning will affect the Yukawa 
coupling matrices. 

In order to avoid fine-tuning, and to give the matrices a non-trivial 
structure, we will assume that the matrices will have a Fritzsch 
structure\cite{fritzsch}.   The original Fritzsch matrix was of the
form
\begin{equation}
    \pmatrix{0&A&0\cr A&0&B\cr 0&B&C\cr}
    \end{equation}
    where $C \sim  m_{\tau}$, $B \sim \sqrt{m_{\mu}m_{\tau}}$ and 
    $A \sim \sqrt{m_{e}m_{\mu}}$.  This matrix has the correct 
    eigenvalues, is parameter-free and does not have fine-tuning.  It 
    was shown in Ref. \cite{chengsher} that 
a wide variety of matrices, such as those with nonzero values in the 
$1,1$ and $2,2$ elements, will (if one requires that there be no 
fine-tuning) yield the same flavor-changing-neutral structure as the 
Fritzsch structure.   We expect that the general case will give the 
same qualitative results.

Since the matrices we are considering are not symmetric,
we will write the desired mass matrix as
\begin{equation}
\pmatrix{0 & a\sqrt{m_{e}m_{\mu}}& 0 \cr b\sqrt{m_{e}m_{\mu}} & 0 & c\sqrt{m_{\mu}m_{\tau}}\cr 0 & d\sqrt{m_{\mu}m_{\tau}} & e m_{\tau}\cr}
\end{equation}
where $a,b,c,d$ and $e$ are all of order $1$.  In general, with 
multiple scalars, the individual Yukawa couplings would be of 
this form, with $\sum a=\sum b= \sum c = \sum d = \sum e = 1$.

So, for a given model, and a given choice of permutations of $i,j$ 
and $k$, one compares this matrix with the mass matrices $M_{A}$ and 
$M_{B}$, and reads off the values of $a,b,c,d$ and $e$.  Then the mass 
matrices are diagonalized, and the Yukawa coupling matrices in the 
diagonal basis are determined.  It turns out that the procedure is 
only consistent for Model A if $j$ is the second generation, and thus 
we have a total of 4 Yukawa coupling matrices for Model A (two 
choices of $\Phi_{1}$ or $\Phi_{2}$ , and the choice between 
$i=1,k=3$ or $i=3,k=1$), and 12 Yukawa coupling matrices for Model B
(two choices of $\Phi$ and six permutations of $i,j,k$).  However, the 
results are simplified in Model B by the fact that if we 
permute the first and third indices, the Yukawa coupling matrices are 
identical, so there are only six different matrices.

The Yukawa couplings are given in Table 2 for Model A and in Table 
3 for Model B.  We label Model A1 and A2 as corresponding to
$(i,j,k) = (e,\mu,\tau)$ or $(\tau,\mu,e)$, respectively, and we 
label Model B1, B2 and B3 as corresponding to 
$(e,\mu,\tau), (e,\tau,\mu)$ or $(\mu,e,\tau)$, respectively.
\begin{table}[h]
\begin{tabular}{|c|c|c|}
    \hline scalar&A1&A2 \\ \hline $\Phi_{1}$ & $\pmatrix{0 & -\sqrt{m_{e}m_{\mu}}& -m_{\mu}\sqrt{m_{e}\over m_{\tau}}\cr
     0 & -m_{\mu} & -m_{\mu}\sqrt{m_{\mu}\over m_{\tau}}\cr \sqrt{m_{e}m_{\tau}} & \sqrt{m_{\mu}m_{\tau}} & m_{\mu}}$ &
      $\pmatrix{0 & 0 & \sqrt{m_{e}m_{\tau}}\cr -\sqrt{m_{e}m_{\mu}} & -m_{\mu} & \sqrt{m_{\mu}m_{\tau}}\cr
       -m_{\mu}\sqrt{m_{e}\over m_{\tau}} & -m_{\mu}\sqrt{m_{\mu}\over m_{\tau}} & m_{\mu}}$ \\ \hline $\Phi_{2}$ &
        $\pmatrix{m_{e} & \sqrt{m_{e}m_{\mu}}& m_{\mu}\sqrt{m_{e}\over m_{\tau}}\cr 0 & 0 & 0\cr -\sqrt{m_{e}m_{\tau}} &
	 -\sqrt{m_{\mu}m_{\tau}} & m_{\tau}+m_{\mu}}$ & $\pmatrix{m_{e} & 0 & -\sqrt{m_{e}m_{\tau}}\cr \sqrt{m_{e}m_{\mu}} &
	  0 &-\sqrt{m_{\mu}m_{\tau}}\cr m_{\mu}\sqrt{m_{e}\over m_{\tau}} & 0 & m_{\tau}+m_{\mu}}$ \\  \hline
    \end{tabular}
\caption{Yukawa coupling matrices to $\Phi_{1}$ and $\Phi_{2}$ for
Model A.  All entries are to be divided by
$\sqrt{v^{2}_{1}+v^{2}_{2}}/\sqrt{2}= 175 {\rm\ GeV}$. The specific models are discussed in the text.}
\end{table}

\begin{table}[ht]
\begin{tabular}{|c|c|c|}
    \hline scalar&B1&B2\\ \hline 
    $\Phi_{1}$ & $\pmatrix{0&-\sqrt{m_{e}m_{\mu}}&\sqrt{m_{e}m_{\tau}}\cr 0 & -m_{\mu} &
    \sqrt{m_{\mu}m_{\tau}}\cr 0 & -m_{\mu}\sqrt{m_{\mu}\over m_{\tau}} & m_{\mu}\cr}$ & 
    $\pmatrix{0&0&-\sqrt{m_{e}m_{\tau}}\cr 0&0&-\sqrt{m_{\mu}m_{\tau}}\cr 0&0&m_{\tau}+m_{\mu}\cr}$ \\ \hline 
    $\Phi_{2}$ & $\pmatrix{m_{e}& \sqrt{m_{e}m_{\mu}}&-\sqrt{m_{e}m_{\tau}}\cr 
    0&0&-\sqrt{m_{\mu}m_{\tau}}\cr
    0&0&  m_{\tau}+m_{\mu}\cr} 
    $ & $\pmatrix{m_{e}&0&\sqrt{m_{e}m_{\tau}}\cr 0&-m_{\mu}& \sqrt{m_{\mu}m_{\tau}}\cr
     0 & -m_{\mu}\sqrt{m_{\mu}\over m_{\tau}}&m_{\mu}\cr}$\\ \hline
    \end{tabular}
    
    \vskip 0.5cm
    \begin{tabular}{|c|c|} \hline scalar&B3 \\ \hline $\Phi_{1}$ &
      $\pmatrix{m_{e}&\sqrt{m_{e}m_{\mu}}&m_{\mu}\sqrt{m_{e}\over m_{\tau}}\cr 0&0&0\cr 0&0&0\cr}$ \\ \hline 
      $\Phi_{2}$  &$\pmatrix{0&\sqrt{m_{e}m_{\mu}}&-m_{\mu} \sqrt{m_{e}\over m_{\tau}}\cr 
      0&-m_{\mu}& -m_{\mu}\sqrt{m_{\mu}\over m_{\tau}}\cr 0& -m_{\mu}\sqrt{m_{\mu}\over m_{\tau}}& m_{\tau}+2m_{\mu}\cr}$
       \\ \hline
       \end{tabular}
\caption{ Yukawa coupling matrices to $\Phi_{1}$ and $\Phi_{2}$ for
Model B.  All entries are to be divided by
$\sqrt{v^{2}_{1}+v^{2}_{2}}/\sqrt{2}= 175 {\rm\ GeV}$. The specific models are discussed in the text.}
\end{table}
Note that we have tacitly assumed that the two Higgs
triplets in the low-energy sector do not mix.  This is for
simplicity.  One can easily find the couplings of one of the physical Higgs 
bosons by including an appropriate (and unknown) mixing angle.   In our 
discussion of the phenomenology, 
this angle will play an important role, and it must be kept in mind.
 
Note how unusual some of these Yukawa coupling matrices are.  For 
example,  in Model B3's coupling to $\Phi_{1}$, the Yukawa couplings to $\tau-\tau$, $\mu-\tau$ 
and $\mu-\mu$ all vanish, leading to an effectively leptophobic Higgs 
boson.   We now turn to the lepton-flavor-changing 
phenomenology of these models.

\section{Leptonic Flavor-changing Decays}

In all of these models, there are Higgs-mediated lepton-flavor-changing
neutral currents (FCNC) arising from the off-diagonal terms in the Yukawa coupling matrices.
 This will lead to $\mu$ and $\tau$ decays which
violate lepton number.   The leptonic decays of the $\tau^{-}$ are 
into $e^{-}e^{-}e^{+}$, $\mu^{-}\mu^{-}\mu^{+}$, $e^{-}e^{-}\mu^{+}$,
$\mu^{-}\mu^{-}e^{+}$, $e^{-}\mu^{-}\mu^{+}$, $e^{-}\mu^{-}\mu^{+}$ and
the $\mu$ decay is into $e^{-}e^{-}e^{+}$.    

The decay rate calculations are straightforward\cite{sheryuan, jairo}.  
Given the experimental upper bound on the decay rate for each of 
these processes, one can find a lower bound on the mass of the 
exchanged Higgs boson.  The rate is inversely proportional to the 
Higgs mass to the fourth power.   Examining all of the Yukawa coupling 
matrices in the previous section, we find that this lower bound is 
always less than $4.9$ GeV.  Since the experimental lower bound is 
more than an order of magnitude higher, these bounds are not 
competitive.

One can still have one-loop radiative decays.  Again, the bounds 
from $\tau$ decays ($\tau\rightarrow e\gamma, \tau\rightarrow 
\mu\gamma$) do not give strong bounds.  The strongest is from 
$\tau\rightarrow\mu\gamma$ in Models A1, A2, B1, B2 in which the 
first three involve coupling to $\Phi_{2}$ and the last to
$\Phi_{1}$.  However, even this lower bound is only $50$ GeV, and is 
marginally competitive with current experimental bounds.

A much stronger bound comes from $\mu\rightarrow e\gamma$.  Here a 
$\tau$ can be in the loop.  The formula for the decay 
rate\cite{generalrate} is
\begin{equation}
    \Gamma_{\mu\rightarrow e\gamma}= h^{2}_{\mu\tau}h^{2}_{e\tau}{\alpha m^{2}_{\tau}m_{\mu}^{3}\over 128\pi^{4}} 
    \left[ {\ln(m_{h}/m_{\tau})\over m_{h}^{2}} \right]^{2}
    \end{equation}
where the $h_{ij}$ are the Yukawa couplings, and $m_{h}$ is the 
scalar mass.  This result does not change if the relevant scalar is a
pseudoscalar.

Plugging in, one finds a lower bound of $230$ GeV on the 
exchanged scalar mass for models A1, A2, B1 and B2, regardless of 
which scalar is used.  However, for several reasons this bound is quite uncertain.
  First, we have a Fritzsch ansatz, and without
that assumption the Yukawa couplings are only order-of-magnitude.
Second, we have 
ignored mixing angles, which could also lower the Yukawa couplings 
substantially.  Third, these 
models can have heavy leptons in the loop, and cancellations are 
possible.  Thus, the numerical bound should be taken {\it cum grano 
salis}, but it is clear that $\mu\rightarrow e\gamma$ may be 
quite close to detection in these models.

Note that Model B3 was not included in the above paragraph.  In the 
coupling to $\Phi_{1}$, there is no bound coming from muon decay; in 
the coupling to $\Phi_{2}$, there is a bound of $7.3$ GeV on the Higgs 
mass.  So the model is unconstrained by muon decay, and 
the Higgs bosons in this model could be very light.  

We now turn to lepton-number violation in Higgs decays.

\section{Lepton-number violating Higgs Decays}

We have a two-Higgs model in the low-energy sector.   Here, mixing 
between the Higgs scalars (which will generically occur and depend on 
parameters of the scalar potential) can have a major effect on the 
branching ratios of Higgs bosons.  For the moment, we will ignore 
these effects, but they are important and will be discussed shortly.

In the conventional two-Higgs model, one Higgs doublet
couples to the $Q=2/3$ quarks, and the other to the $Q=-1/3$ quarks 
and the charged leptons.  The latter's primary decay into fermions is 
thus to $b\overline{b}$, with the $\tau^{+}\tau^{-}$ decay being a 
factor of ${3 m_b^2 \over m_\tau^2} \sim 25$ smaller.  Of course, the primary decay mode could be
$WW, WW^{*},ZZ $ or $ZZ^{*}$, depending on the mass of the Higgs.  Here we will 
only look at the primary fermionic decays, which are relevant if
the Higgs mass is not too much larger than its current lower bound (if 
it is larger, the fermionic decay branching ratios might be small, but 
certainly detectable at the LHC).   The primary fermionic decay mode of the 
Higgs that couples to $Q=2/3$ fields would be into $t\overline{t}$ if 
kinematically accessible, and $c\overline{c}$ if not.  It will not
couple to the charged leptons.  If the mixing 
angle is not too small, then the latter field's primary fermionic decay 
is also into $b\overline{b}$.

In both models under consideration, one of the quark generations has 
a different structure than the other two.  The unique generation is 
generally assumed to be the third generation, an assumption we concur with.
  If it is not, there will be flavor-changing effects in the kaon sector which will be
phenomenologically problematic.

Then, again ignoring mixing, the scalar that couples to 
$b\overline{b}$ will {\it not} couple to the charged leptons.
The field coupling to the charged leptons will couple to the strange 
quark and to the top quark.   If its mass is below $360$ GeV, then 
its primary fermionic decay is into the charged leptons and the 
strange quark\footnote{Actually, if it between $270$ and $360$ GeV, 
then the three body decay through a virtual top into $tbW$ will 
dominate.}.  In this case, we can calculate the fermionic branching ratios for 
the five models under consideration, and show these in Table 4.
\begin{table}[h]
\begin{center}
\begin{tabular}{|c|c|c|c|c|}\hline
    Model&$\mu\mu$&$\mu\tau$&$\tau\tau$&$s\overline{s} $\\ 
    \hline\hline
    A1&0&.05&.94&.01\\
    A2&0&.06&.93&.01\\
    B1&.04&.72&.04&.20\\
    B2&0&.06&.93&.01\\
    B3&0&0&0&100\\ \hline
    \end{tabular}
\end{center}
\caption{The fermionic branching fraction into various final states
for the Higgs that does not couple to the b-quarks in the various 
models.  We have explicitly assumed no mixing between the Higgs 
scalars, and that top quark decays are not kinematically accessible.
The decay into gauge bosons will dominate if they are 
kinematically accessible.}
\end{table}

The results in Table 4 are interesting.  In Models A1, A2 and B2, we
see that the inversion of the bottom-top quark doublet takes the field 
that would ``normally'' decay into $b\overline{b}$, and (since the top 
quark is too heavy) makes its primary decay mode $\tau^{+}\tau^{-}$.  
This would be a very dramatic signature.   In Model B3, in which the 
Higgs is leptophobic (and in which, as shown in the last section, 
radiative muon decay does not bound the Higgs mass), there are no leptonic decays, and the primary 
decay mode would be into $s\overline{s}$.   The most unusual model is 
B1, in which the {\it primary} decay mode is into $\mu\tau$.  This 
monochromatic muon would give a very dramatic signature.

All of these signatures are quite dramatic.  How realistic is this 
scenario?  Abandoning the use of the Fritzsch ansatz will have effects of $O(1)$ 
on these results, but will not change the general results.  However, 
the assumption of no mixing between the doublets will have a 
substantial effect on the scalars (the pseudoscalar will not, in
general, have this mixing, and thus the results of the above 
paragraph will apply).  For the scalars, mixing means that the
branching ratio into $b\overline{b}$ is not negligible.  For 
Models A1, A2 and B2, the fermionic branching ratio into $b\overline{b}$
relative to $\tau^{+}\tau^{-}$ is approximately $25\sin^{2}\theta$, 
and thus the individual branching ratios must be reduced 
accordingly.  For Model B3, the fermionic branching ratio into
$b\overline{b}$ is approximately $1000\sin^{2}\theta$, and thus the 
primary decay mode will almost certainly be into $b\overline{b}$, 
unless the angle is extremely small.  For B1, the fermionic branching 
ratio into $b\overline{b}$ is approximately $400\sin^{2}\theta$, and 
thus it is likely that $b\overline{b}$ decays will dominate, although 
the remarkable $\mu\tau$ decay mode will still be substantial.  Note 
that the signature for $\mu\tau$ decays is very clean, and branching 
ratios of $10^{-4}$ can be detected.   As a result, in all of these 
models except B3, the Higgs decay into $\mu\tau$ is detectable.

\section{Bounds on the Gauge Boson Sector}

The electroweak Lagrangian (with the kinetic terms dropped) may be written in the form
\begin{displaymath}
{\cal L}=
\sum_i \bar{\psi}_i({g \over 2} \lambda_\alpha A_\alpha^\mu + g'XB^\mu)\psi_i =\sum_i \bar{\psi}_i
 \pmatrix{ D_1^\mu & {g \over \sqrt{2} } W^{+\mu} & {g \over \sqrt{2} } K^{+\mu} \cr
          {g \over \sqrt{2} } W^{-\mu} & D_2^\mu & {g \over \sqrt{2} } K^{0\mu} \cr
	  {g \over \sqrt{2} } K^{-\mu} & {g \over \sqrt{2} } \bar{K}^{0\mu} & D_3^\mu \cr}\psi_i
\end{displaymath}
where
\begin{eqnarray}
D_1^\mu=g \left( {A_3^\mu \over 2} + {A_8^\mu \over 2\sqrt{3} } \right) +g'XB^\mu \nonumber \\
D_2^\mu=g \left( -{A_3^\mu \over 2} + {A_8^\mu \over 2\sqrt{3} } \right) +g'XB^\mu \nonumber \\
D_3^\mu= -g{A_8^\mu \over \sqrt{3} } +g'XB^\mu
\end{eqnarray}
and the sum is over all $\psi$ in the model.
With the relationship $\sin^{2}\theta_W={3g'^2 \over 3g^2+4g'^2}$ defining the electroweak mixing angle, we find that
 the diagonal terms reduce to combinations of the expected neutral gauge bosons $A^\mu$ and $Z^\mu$, plus a new boson,
  the $Z'^\mu$.
The photon and $Z$ have the same couplings and Feynman rules as the SM, and therefore display no unusual characteristics.
However, the $Z'$ has vector and axial couplings which depend on the particular lepton generation, Eqs.\ref{eq:L1}-\ref{eq:L4},
 leading to FCNC.

In terms of the $SU(3)_L \otimes U(1)_X$ gauge bosons, we find that the low-energy fields are given by~\cite{diaz, ochoa, VanDong:2005pi}
\begin{eqnarray}
A_\mu & = & S_W A_\mu^3 + C_W \left( {T_W \over \sqrt{3} } A_\mu^8 +\sqrt{ 1-{T_W^2 \over 3}} B_\mu \right) \cr \nonumber
Z_\mu & = & C_W A_\mu^3 - S_W \left( {T_W \over \sqrt{3} } A_\mu^8 +\sqrt{ 1-{T_W^2 \over 3}} B_\mu \right) \cr \nonumber
Z_\mu' & = & -\sqrt{ 1-{T_W^2 \over 3}} A_\mu^8 + {T_W \over \sqrt{3} } B_\mu ,
\label{eq:bosons}
\end{eqnarray}
where $S_W=\sin \theta_W$, $C_W=\cos \theta_W$, and $T_W=\tan \theta_W$.
These fields have the eigenvalues
\begin{equation}
M^2_{A_\mu}=0; \quad M^2_{Z_\mu}\simeq {g^2 \over 2} \left[ { 3g^2 +4g'^2 \over 3g^2 + g'^2} \right] (v_1^2 +v_2^2);
 \quad M^2_{Z_\mu'}\simeq {2 [3g^2 +g'^2] \over 9} V^2.
\end{equation}

The $Z'$ has a vertex factor of the form $-i {1 \over 2} \gamma_\mu \left( C_V - C_A \gamma_5 \right)$ where
 the $C_{V,A}$ are family-dependent, and given in Table 5.
\begin{table}[h]
\begin{center}
\begin{tabular}
{ccc}\hline \hline
Family & $C_V$ & $C_A$ \\ \hline
$L_1$,$L_4$ & ${10 S_W \over \sqrt{3-4S_W^2}} + {\sqrt{3-4S_W^2} \over S_W}$ &  ${\sqrt{3-4S_W^2} \over S_W} - {2 S_W \over \sqrt{3-4S_W^2}}$ \\
$L_2$ & ${8 S_W \over \sqrt{3-4S_W^2} }-{\sqrt{3-4S_W^2} \over S_W}$ & $-{\sqrt{3-4S_W^2} \over S_W} - {4 S_W \over \sqrt{3-4S_W^2} }$ \\
$L_3$ & ${6 S_W \over \sqrt{3-4S_W^2} } -3{\sqrt{3-4S_W^2} \over S_W}$ & ${\sqrt{3-4S_W^2} \over S_W} - {2 S_W \over \sqrt{3-4S_W^2}}$ \\ \hline \hline
\end{tabular}
\end{center}
\caption{The $C_V$ and $C_A$ for the various lepton families. 
A common factor of ${e\over 6\cos\theta_{W}}$ has been factored out of each.
Note that $C_A$ is the same for $L_{1,3,4}$.}
\end{table}

A recent analysis of precision electroweak (EW) bounds in 331 models
without exotic electric charges\cite{ochoa} gave a lower bound of
$1400$ GeV on the mass of the $Z^{\prime}$.
Since the $SU(3)_L \otimes U(1)_X$ representations are different for each lepton family, one expects $Z'$-mediated FCNC.
As discussed in the last section, the mixing matrix between the $SU(3)_L$ eigenstates and the mass eigenstates will
have too many free parameters.
To estimate the size of the $Z'$ FCNC, we therefore again use the Fritsch ansatz.
Failure to use this ansatz results in too many parameters.
This results in a mixing matrix with no free parameters but the lepton masses.
To determine the FCNC couplings of the $Z'$, one picks the model and diagonalizes.
Using the $C_V$ and $C_A$ in Table 5, one reads off the couplings for each particle.
These couplings will be a linear combination of the family couplings.
Since the $C_V$ differ for each family, there will be FCNC.

The most stringent bound on $M_{Z'}$ is found from $\mu \rightarrow 3e$ decays.
The formula for this decay rate is
\begin{equation}
\Gamma = {\pi m_\mu^5 \over 108} \left({ e \over 24 \pi C_W M_{Z'}} \right)^4
\left[ 3(C_{Ve\mu}^2 + C_{Ae\mu}^2 )(C_{Vee}^2 + C_{Aee}^2 )
 +4C_{Ve\mu}C_{Ae\mu}C_{Vee}C_{Aee} \right]
\end{equation}
Given that we do not know which family corresponds to which lepton, we try all possibilities.
This provides bounds that range from $2$ TeV in Model B2, to between 20 and 40 TeV in the other models.
A similar calculation using $\tau \rightarrow 3\mu$ or  $\mu \rightarrow e \gamma$ provides much weaker lower bounds.
Thus precision EW bounds will not be relevant in these models.

A bound of 20 to 40 TeV is discouraging since the $Z'$ will be beyond the reach of the LHC, and because fine-tuning
will be needed to explain a new hierarchy problem. 
Nonetheless, Model B2 does not need substantial fine-tuning, and the Higgs decays in any of the models will provide
distinct signatures.

\section{Conclusions}

We have studied a pair of 3-3-1 models that have not previously been examined.
The defining characteristic of these models is that each lepton generation has a unique structure.
This leads to FCNC decays mediated by the light Higgs and $Z'$ boson.
$Z'$ mediated $\mu \rightarrow 3e$ provides a lower bound of $2$ TeV for $M_Z'$ in Model B2, and between 20 and 40 TeV in
the others.
These models will all have interesting Higgs decay signatures.
In particular, $\Phi \rightarrow \mu \tau$ could show up clearly at the LHC.

This research was supported by the
National Science Foundation grant PHY-023400. 

\hspace{0.2in}
\appendix
\noindent \bf \Large Appendix A
\normalsize \rm \\

Here are the full mass matrices for the charged leptons in the 3-3-1 models studied in this paper.
For Model A we have:
\begin{equation}
\pmatrix{ h_1 v_2 & h_2 v_2 & 0 & h_3 v_2 & 0 \cr
h_7 v_1 & h_8 v_1 & -g_1 v_2 & h_9 v_1 & g_2 V \cr
h_{10} v_1 & h_{11} v_1 & -g_3 v_2 & h_{12} v_1 & g_4 V \cr
h_4 V & h_5 V & 0 & h_6 V & 0 \cr
h_{13} v_1 & h_{14} v_1 & -g_5 v_2 & h_{15} v_1 & g_6 V \cr}
\end{equation}
where the ordering is $e_i$, $e_j$, $e_k$, $E_i$, $E_k$.

The relevant terms in the lagrangian are
\begin{eqnarray}
{\cal L}_{Y,AA} &=& \left(h_4 \bar{\psi}_{iL} e_{iR} +
 h_5 \bar{\psi}_{iL} e_{jR} + h_6 \bar{\psi}_{iL} E_{iR}\right)\Phi_A \\ \nonumber
&& + \epsilon_{\alpha \beta \gamma} \left(g_2 \bar{\psi}_{jL}^\alpha (\psi^{\prime\prime c}_{kL})^\beta +
 g_4 \bar{\psi}_{kL}^\alpha (\psi^{\prime\prime c}_{kL})^\beta +
 g_6 \bar{\psi^{\prime}}_{kL}^\alpha (\psi^{\prime\prime c}_{kL})^\beta \right) \Phi_A^\gamma
\end{eqnarray}

\begin{eqnarray}
{\cal L}_{Y,A1} &=& \left( h_7 \bar{\psi}_{jL} e_{iR} +
 h_8 \bar{\psi}_{jL} e_{jR} + h_9 \bar{\psi}_{jL} E_{iR} +
 h_{10} \bar{\psi}_{kL} e_{iR} +
 h_{11} \bar{\psi}_{kL} e_{jR} + h_{12} \bar{\psi}_{kL} E_{iR} \right. \\ \nonumber
&& \left. + h_{13} \bar{\psi^{\prime}}_{kL} e_{iR} +
 h_{14} \bar{\psi^{\prime}}_{kL} e_{jR} + h_{15} \bar{\psi^{\prime}}_{kL} E_{iR} \right) \Phi^*_1
\end{eqnarray}

\begin{eqnarray}
{\cal L}_{Y,A2} &=& \left( h_1 \bar{\psi}_{iL} e_{iR} +
 h_2 \bar{\psi}_{iL} e_{jR} + h_3 \bar{\psi}_{iL} E_{iR} \right) \Phi_2  \\ \nonumber
&& + \epsilon_{\alpha \beta \gamma} \left(g_1 \bar{\psi}_{jL}^\alpha (\psi^{\prime\prime c}_{kL})^\beta +
 g_3 \bar{\psi}_{kL}^\alpha (\psi^{\prime\prime c}_{kL})^\beta +
 g_5 \bar{\psi^{\prime}}_{kL}^\alpha (\psi^{\prime\prime c}_{kL})^\beta \right)\Phi_2^\gamma
\end{eqnarray}

Similarly, the mass matrix for Model B is
\begin{equation}
\pmatrix{ h_1 v_2 & h_2 v_2 & h_3 v_2 & h_4 v_2 & h_5 v_2 & g_4 V & h_6 v_2 \cr
h_{13} v_1 & h_{14} v_1 & h_{15} v_1 & h_{16} v_1 & h_{17} v_1 & 0 & h_{18} v_1 \cr
h_{19} v_2 & h_{20} v_2 & h_{21} v_2 & h_{22} v_2 & h_{23} v_2 & g_5 V & h_{24} v_2 \cr
h_7 V & h_8 V & h_9 V & h_{10} V & h_{11} V & -g_1 v_2 & h_{12} V \cr
h_{25} V & h_{26} V & h_{27} V & h_{28} V & h_{29} V & -g_2 v_2 & h_{30} V \cr
h_{31} v_2 & h_{32} v_2 & h_{33} v_2 & h_{34} v_2 & h_{35} v_2 & g_6 V & h_{36} v_2 \cr
h_{37} V & h_{38} V & h_{39} V & h_{40} V & h_{41} V & -g_3 v_2 & h_{42} V \cr}
\end{equation}
with the ordering $e_i$, $e_j$, $e_k$, $E_i$, $E_{1k}$, $E_{2k}$, $E_{3k}$.

The relevant terms in the Lagrangian are
\begin{eqnarray}
{\cal L}_{Y,BA}&=&\left( h_{7} \bar{\psi}_{iL} e_{iR} +
 h_{8} \bar{\psi}_{iL} e_{jR} + h_{9} \bar{\psi}_{iL} e_{kR} +
 h_{10} \bar{\psi}_{iL} E_{iR} + h_{11} \bar{\psi}_{iL} E_{1kR} +
 h_{12} \bar{\psi}_{iL} E_{3kR} \right. \\ \nonumber
&& + h_{25} \bar{\psi}_{kL} e_{iR} +
 h_{26} \bar{\psi}_{kL} e_{jR} + h_{27} \bar{\psi}_{kL} e_{kR} +
 h_{28} \bar{\psi}_{kL} E_{iR} + h_{29} \bar{\psi}_{kL} E_{1kR} +
 h_{30} \bar{\psi}_{kL} E_{3kR}  \\ \nonumber
&& + h_{37} \bar{\psi^{\prime\prime}}_{kL} e_{iR} +
 h_{38} \bar{\psi^{\prime\prime}}_{kL} e_{jR} + h_{39} \bar{\psi^{\prime\prime}}_{kL} e_{kR} \\ \nonumber
 && \left. + h_{40} \bar{\psi^{\prime\prime}}_{kL} E_{iR} + h_{41} \bar{\psi^{\prime\prime}}_{kL} E_{1kR} +
 h_{42} \bar{\psi^{\prime\prime}}_{kL} E_{3kR} \right) \Phi_A  \\ \nonumber
&& + \epsilon_{\alpha \beta \gamma} \left( g_4 \bar{\psi^{\prime}}_{kL}^\alpha (\psi^{c}_{iL})^\beta +
 g_5 \bar{\psi^{\prime}}_{kL}^\alpha (\psi^{c}_{kL})^\beta +
 g_6 \bar{\psi^{\prime}}_{kL}^\alpha (\psi^{\prime\prime c}_{kL})^\beta \right) (\Phi^*_A)^\gamma
\end{eqnarray}

\begin{eqnarray}
{\cal L}_{Y,B1} &=& \left( h_{13} \bar{\psi}_{jL} e_{iR} +
 h_{14} \bar{\psi}_{jL} e_{jR} + h_{15} \bar{\psi}_{jL} e_{kR} \right. \\ \nonumber
 && \left. + h_{16} \bar{\psi}_{jL} E_{iR} + h_{17} \bar{\psi}_{jL} E_{1kR} +
 h_{18} \bar{\psi}_{jL} E_{3kR} \right)\Phi^*_1
\end{eqnarray}

\begin{eqnarray}
{\cal L}_{Y,B2} &=& \left( h_{1} \bar{\psi}_{iL} e_{iR} +
 h_{2} \bar{\psi}_{iL} e_{jR} + h_{3} \bar{\psi}_{iL} e_{kR} +
 h_{4} \bar{\psi}_{iL} E_{iR} + h_{5} \bar{\psi}_{iL} E_{1kR} +
 h_{6} \bar{\psi}_{iL} E_{3kR} \right. \\ \nonumber
&& + h_{19} \bar{\psi}_{kL} e_{iR} +
 h_{20} \bar{\psi}_{kL} e_{jR} + h_{21} \bar{\psi}_{kL} e_{kR} +
 h_{22} \bar{\psi}_{kL} E_{iR} + h_{23} \bar{\psi}_{kL} E_{1kR} +
 h_{24} \bar{\psi}_{kL} E_{3kR} \\ \nonumber
&& + h_{31} \bar{\psi^{\prime\prime}}_{kL} e_{iR} +
 h_{32} \bar{\psi^{\prime\prime}}_{kL} e_{jR} + h_{33} \bar{\psi^{\prime\prime}}_{kL} e_{kR} \\ \nonumber
 && \left. + h_{34} \bar{\psi^{\prime\prime}}_{kL} E_{iR} + h_{35} \bar{\psi^{\prime\prime}}_{kL} E_{1kR} +
 h_{36} \bar{\psi^{\prime\prime}}_{kL} E_{3kR} \right) \Phi_2 \\ \nonumber
&& + \epsilon_{\alpha \beta \gamma} \left( g_1 \bar{\psi^{\prime}}_{kL}^\alpha (\psi^{c}_{iL})^\beta +
 g_2 \bar{\psi^{\prime}}_{kL}^\alpha (\psi^{c}_{kL})^\beta +
 g_3 \bar{\psi^{\prime}}_{kL}^\alpha (\psi^{\prime\prime c}_{kL})^\beta \right) (\Phi^*_2)^\gamma
\end{eqnarray}

\end{document}